Neural correlates of perceptual consciousness from within: a narrative review of human intracranial research

Francois Stockart[1,2], Alexis Robin[3], Hal Blumenfeld[1,4,5,6], Milan Brazdil[7,8], Philippe Kahane[3], Liad Mudrik[9,10,11], Jasmine Thum[12], Michael Pereira[13]*, Nathan Faivre[2]*

1. Department of Neurology, Yale University, New Haven, CT 06520, USA
2. Univ. Grenoble Alpes, Univ. Savoie Mont Blanc, CNRS, LPNC, 38000 Grenoble, France
3. Neurology Department, CHU Grenoble Alpes, INSERM, U1216, Grenoble Institut Neurosciences, Grenoble, France
4. Interdepartmental Neuroscience Program, Yale University, New Haven, CT 06520, USA
5. Department of Neurosurgery, Yale University, New Haven CT 06520, USA
6. Department of Neuroscience, Yale University, New Haven, CT 06520, USA
7. 1st Department of Neurology, St. Anne Univ. Hospital and Faculty of Medicine, Masaryk University, Brno, Czech Republic
8. Behavioral and Social Neuroscience Research Group, CEITEC MU, Brno, Czech Republic
9. School of Psychological Sciences, Tel Aviv University, Tel Aviv, Israel
10. Sagol School of Neuroscience, Tel Aviv University, Tel Aviv, Israel
11. Brain, Mind, and Consciousness Program, Canadian Institute for Advanced Research (CIFAR), Toronto, ON, Canada
12. Department of Neurosurgery, The University of Alabama in Birmingham, Birmingham, AL, USA
13. Univ. Grenoble Alpes, Inserm, U1216, Grenoble Institut Neurosciences, Grenoble, France

*Shared contribution

## Abstract

Despite many years of research, the quest to identify neural correlates of perceptual consciousness (NCC) remains unresolved. One major obstacle lies in methodological limitations: most studies rely on non-invasive neural measures with limited spatial or temporal resolution making it difficult to disentangle proper NCCs from concurrent cognitive processes. Additionally, the relatively low sensitivity of non-invasive neural measures limits the interpretation of null findings in studies targeting proper NCCs. In this review, we discuss how human intracranial recordings can advance the search for NCCs, by offering high spatiotemporal resolution, improved signal sensitivity, and broad cortical and subcortical coverage. We review studies that have examined NCCs at the level of single neurons and populations of neurons, and evaluate their implications on the debates between cognitive and sensory theories of consciousness. Finally, we highlight the limits of current intracranial human recordings and propose future directions based on emerging technologies and novel experimental paradigms.

## Neural correlates of consciousness

Perceptual consciousness refers to the subjective experience associated with processing sensory stimuli. When presented with images, sounds, touches, tastes, or smells, humans not only register these inputs but also report vivid conscious experiences. Understanding the mechanisms underlying perceptual consciousness has been a focus of research in philosophy and psychology for centuries (James, 1890; LeDoux et al., 2020). A major shift occurred with the proposal by Francis Crick and Christof Koch (1990) to identify *neural correlates of consciousness* (NCCs) - the minimal neural mechanisms jointly sufficient for a particular conscious percept to occur. In empirical research, studies on NCCs relied mostly on the *contrastive approach* (Baars, 1988; Dehaene, 2014), where neural activity is compared between trials where the same stimulus is consciously perceived or not (see **Figure 1**). In the conscious condition, participants report perceiving a critical stimulus. The same critical stimulus is presented in the unconscious condition[1], but participants report not being conscious of it[2], due to various psychophysical tricks including backward masking, binocular rivalry, (continuous) flash suppression, the attentional blink and the presentation of near-threshold stimuli (see Kim & Blake, 2005; Breitmeyer, 2015 for reviews). Participants are asked to make subjective reports of consciousness (Seth et al., 2008), which allows distinguishing between trials in which they did and did not consciously perceive the stimulus (see Fahrenfort et al., 2025, for potential issues with this approach). Crucially, the two conditions must differ minimally in terms of the stimulus presentation parameters so that the observed neural differences can be attributed to perceptual consciousness, rather than differences in display parameters. Apart from the contrastive approach, recent studies have also relied on a supraliminal approach, in which participants are presented with clear but task-irrelevant stimuli (see Cogitate Consortium et al., 2025; Gerber et al., 2017; Noy et al., 2015; **Figure 1** and below for further details).

It is also helpful to distinguish between levels of consciousness, defined as a global level of arousal or wakefulness (e.g., being awake vs. under anesthesia), and the contents of consciousness, defined as the specific subjective experiences one has while conscious (e.g., perceiving a visual stimulus; Bayne et al., 2016; Laureys, 2005). Though the majority of this review focuses on the contents of consciousness, the two dimensions are intrinsically linked, as global states typically set the conditions for the occurrence of specific conscious contents.

---

[1] In contrastive studies that use backward masking to render stimuli invisible (e.g., Dehaene et al., 1998; 2001), while the critical stimulus is the same across all trials, the mask is presented earlier relative to onset of the critical stimulus or for a longer duration in the unconscious than the conscious condition. This systematic difference in stimulation parameters across conditions is a possible confound when interpreting neural activity. In recent years, the field has moved to paradigms where stimulation parameters are the same in both conditions (e.g., near-threshold paradigm depicted in **Figure 1**).

[2] Note that NCC research differs from the study of the neural correlates of unconscious processing, which is focused on the processing of stimuli that participants do not consciously perceive (Mudrik & Deouell, 2022).

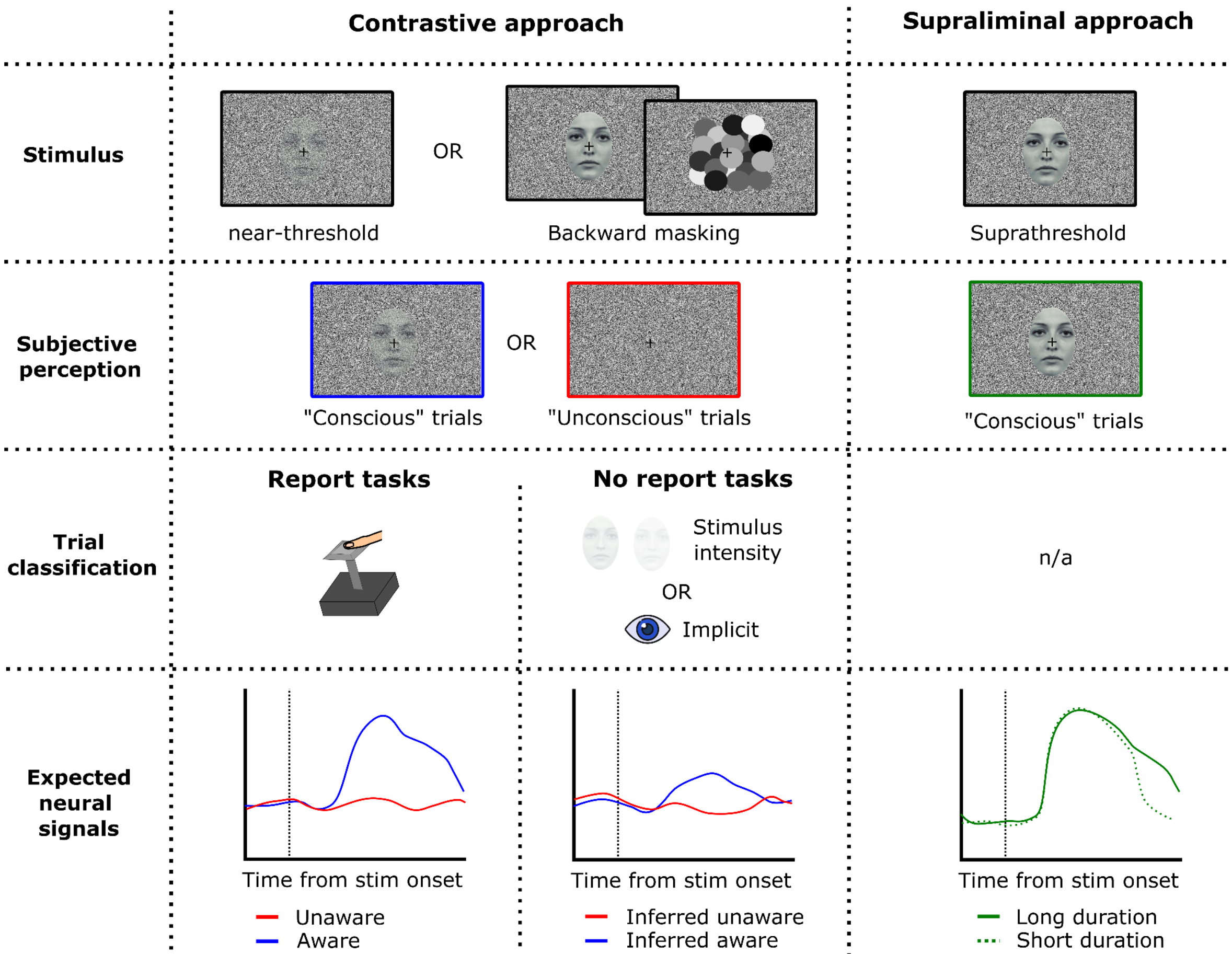

*Figure 1.* Experimental approaches to study the NCC. In the contrastive approach (left), the visibility of stimuli is manipulated across trials such that they are either perceived ("conscious trials") or not perceived ("unconscious trials"). Two of the most popular paradigms to manipulate visibility, near-threshold presentation of stimuli and backward masking, are depicted. In report tasks, participants perform button presses or saccadic eye movements to report whether they perceived the stimulus or not. In no-report tasks, attentional manipulations, stimulus intensity or automatic oculomotor or pupillary responses (e.g., optokinetic nystagmus, pupil dilation) are used to infer if the stimulus was perceived or not. Neural signals that correlate with consciousness should differ across trials that are classified as conscious and unconscious. In the supraliminal approach (right), full contrast stimuli are presented and are assumed to always be consciously perceived. Trials are compared across conditions that differ in the characteristics of the stimuli, such as presentation duration (discussed below).

Since the late 1990s, both the contrastive and the supraliminal approaches have resulted in a multitude of NCC candidates, mostly identified through noninvasive neuroimaging methods (for reviews, see Boly et al., 2017; Dykstra et al., 2017; Koch et al., 2016; Rees et al., 2002). Numerous anatomical maps of the NCCs have been developed over the past thirty years, but no consensus has been reached even on basic properties, such as whether the NCCs include prefrontal regions or not (Boly et al., 2017; Odegaard et al., 2017). The reason for this situation is threefold. In what follows, we discuss the key aspects of these three issues before presenting human intracranial recordings as a potential solution to advance our understanding of perceptual consciousness.

*Distinguishing NCCs from their prerequisites and consequences*

The traditional use of the contrastive approach has important pitfalls (Aru et al., 2012a; De Graaf et al., 2012; Sergent & Naccache, 2012), which are also relevant to intracranial research. First, the brain processes that precede conscious experience differ systematically between conscious and unconscious trials due to various factors including arousal, attention, and expectations. Thus, most studies cannot distinguish a proper NCC from preceding processes. Second, post-perceptual activity following conscious experience can be mistaken for the NCC. For example, because participants are asked to report their perception of the stimuli, the neural activity measured in the unconscious vs. conscious conditions is contaminated by the planning and execution of these reports. We define a report as any explicit behavioral response (whether verbal, manual, or otherwise) that communicates a participant's subjective state. We consider proper NCCs to consist of neural activity that reflects the subjective experience itself, regardless of the downstream requirements of report. Identifying proper NCCs thus requires experimental approaches that separate perceptual experience from its neural precursors and consequences (Aru et al., 2012a; De Graaf et al., 2012; Sergent & Naccache, 2012).

Much research has focused on addressing the consequences problem, notably through *no-report paradigms*, where participants passively view stimuli while their brain activity is recorded (Frässle et al., 2014; Lumer et al., 1998; Tsuchiya et al., 2015a). The rationale is that such paradigms yield recordings that are not confounded by much of the post-perceptual activity associated with reporting one's perception. Three strategies have been employed to infer conscious perception without reports. The first relies on inattentional blindness (Rock et al., 1992). In a first phase, participants perform a task on central stimuli while critical stimuli are presented in the periphery unbeknownst to them. Subsequently, they are informed about the presence of the critical stimuli, which remain task-irrelevant (phase 2) or become central to the task (phase 3). Neural activity in response to the critical stimuli can then be compared between trials where they were consciously perceived (phases 2 and 3), and trials where they were not (phase 1; Dellert et al., 2021; Pitts et al., 2012; 2014). In the second strategy, participants passively observed stimuli, and involuntary eye movements are used as an indirect measure for classifying stimuli as perceived or non-perceived (Frässle et al., 2014; Hesse & Tsao, 2020; Kronemer et al., 2022; White et al., 2022). The third strategy involves presenting stimuli at varying physical intensities, either below or above participants' perceptual threshold (Pereira et al., 2021; Sergent et al., 2021). Researchers can then compare neural responses across these stimulus intensities, assuming that participants are more likely to perceive the stimuli when presented at a higher intensity. The resulting comparison cannot be interpreted in isolation, as the display parameters are varied across conditions. Nevertheless, it can still be very informative when compared with the outcomes of a report condition in which these parameters remain constant.

No-report paradigms have changed the field's perspective on the NCC. Typically, if a seen vs. unseen effect is found in the report condition, but not in the no-report condition, it is interpreted as reflecting processes related to post-perceptual processing rather than to consciousness per se (Dembski et al., 2021). There has been less research on distinguishing the proper NCCs from their prerequisites. However, this reasoning necessarily involves inference based on null effects that can be difficult to interpret, also due to the limitations of brain imaging techniques (for limitations of noninvasive measures, see the following section).

*Limits of noninvasive measures*

The quest for NCCs coincided with the emergence of new technological advancements in computing power and brain imaging techniques. Notably, functional magnetic resonance imaging studies (fMRI) allowed for whole-brain coverage with good spatial resolution, promising to reveal the neural networks involved in consciousness. NCCs were initially investigated using univariate statistical models (Dehaene et al., 2001; Lumer et al., 1998; Tong et al., 1998), and then with the help of multivariate pattern analysis (Hatamimajoumerd et al., 2022; Haynes, 2009). Because of the low temporal resolution of fMRI, however, the temporal dynamics of NCCs remain largely unexplored in these studies. Other neuroimaging modalities, such as electroencephalography (EEG) and magnetoencephalography (MEG), offer significantly better temporal resolution. EEG and MEG can be paired with the contrastive approach to compare the temporal distribution of event-related responses to consciously perceived vs. non-perceived stimuli (Dehaene et al., 2001; King et al., 2016; Koivisto & Revonsuo, 2003; Sergent et al., 2021). Several candidate NCCs were identified using scalp EEG or MEG, but none of them reached a consensus (for reviews, see Bola & Doradzinska, 2021; Dembski et al., 2021; Förster et al., 2020). This might be related to the anatomical limitations of scalp EEG and MEG signals given the overlap of deep and superficial brain signals at the scalp level. This is the "inverse problem", where different cortical configurations can produce identical scalp topographies. Additionally, the signal-to-noise ratio of these measures is unequal across brain regions: MEG is relatively insensitive to radial sources and both methods to subcortical sources (Piastra et al., 2020). In the case of scalp EEG, the scattering of the adjacent electrical signals through the scalp further reduces signal-to-noise ratio from all sources.

The limitations of these methods become more prominent when dealing with null effects such as those often obtained in no-report paradigms. Subtle effects that are transient or involve only small populations of neurons might remain undetected. One non-invasive tool for bridging across spatial and temporal scales is combined fMRI-EEG, where the two signals are collected concurrently and then merged (e.g., Dellert et al., 2021). While powerful, this approach relies on merging an indirect metabolic signal with a weak electrophysiological one filtered by the skull, which is computationally complex and often noisy. More precise invasive measures, which provide direct measures of both local field potentials and spiking activity, thus offer important complementary evidence in perceptual consciousness research.

*Limits of consciousness research in non-human species*

Using invasive electrophysiological measures, non-human animal studies have shown clear prefrontal contributions in no-report paradigms that were not readily observed with fMRI or scalp EEG (e.g., Kapoor et al., 2022; for reviews, see Block, 2024; Panagiotaropoulos, 2024). While non-human electrophysiology combines unique strengths, such as the ability to collect numerous trials and implant dense electrode arrays in specific brain areas selected for research, it also has limitations. First, the range of available behavioral paradigms is more restricted. Paradigms that directly assess subjective states, like more complex scales to measure consciousness (Ramsøy & Overgaard, 2004), cannot be employed. Relatedly, non-human animals are unable to provide verbal reports of their subjective experiences. Second, non-human primate studies usually draw conclusions from a low number of individuals (typically two), who are extensively trained to perform psychophysical tasks that yield rewards, eliciting strong reinforcement learning. Therefore, conclusions from these studies may be jeopardized by specific confounds that are irrelevant to human research. Third, although mapping the extent and signatures of perceptual consciousness across species is valuable (Lamme, 2022), the extent to which these results apply to human consciousness remains unclear. Finally, invasive animal research presents ethical challenges that should not be overlooked, particularly relevant in consciousness research (Mazor et al., 2023).

Given this situation, along with the limitations of fMRI and scalp EEG in studying the NCC, human intracranial EEG seems to be a particularly promising direction for consciousness research. Below, we report the results of a non-systematic literature review, based on 37 articles published between 2001 and 2025. The articles were identified during the course of our own research on NCCs, and supplemented by a bibliographic search based on screening the citations of two leading early publications in the field (Fisch et al., 2009; Gaillard et al., 2009; search performed on April 16th 2025) as well as screening of all the records using intracranial EEG on the ConTrast database on theories of consciousness (Yaron et al., 2022).

## Human intracranial electrophysiology

Human intracranial electroencephalography is typically conducted on patients with drug-resistant epilepsy, who are implanted with electrodes as part of the pre-surgical evaluation of the cortical source of their seizures (Mercier et al., 2022). Broadly, two types of implants are utilized: (1) Electrocorticography (ECoG), where grids and/or strips of electrodes are placed on the surface of the cortex, and stereo-encephalography (sEEG), which employs depth electrodes (**Figure 2**). Both ECoG and sEEG capture local field potentials, which are considered to reflect the activity of large populations of neurons spanning several $mm^3$. In a processing step called rereferencing, neighboring contacts on an electrode can be referenced to each other to ensure that only local activity is captured by the resulting channel (Li et al., 2018). Single-neuron recordings can also be collected using microwires at the tip of sEEG electrodes or by microelectrode arrays on the cortical surface, which allows testing of finer grained hypotheses about the NCCs (Despouy et al., 2020; Fried et al., 1999).

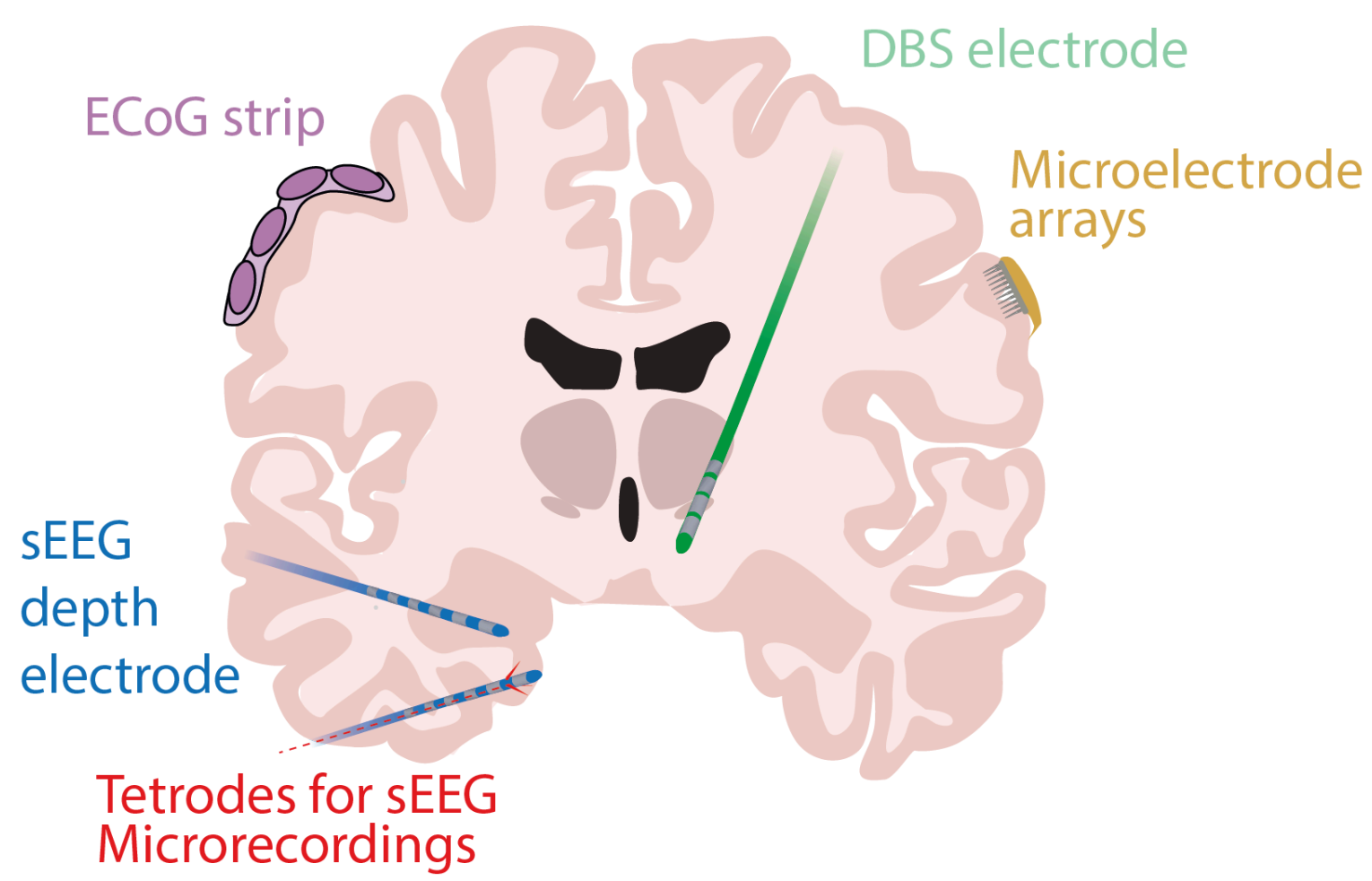

*Figure 2*. Different types of implants used for intracranial recordings in humans: ECoG strips or grids are typically placed under the dura mater (purple implant); Deep-brain stimulation (DBS) electrodes are typically used to record and stimulate subcortical regions (green implant); Microelectrode arrays are positioned on the cortex (yellow implant); sEEG depth electrodes are inserted in the brain across cortical and subcortical structures (blue implant). Hybrid models can record single-unit activity with either tetrodes (red) or microwires protruding from the electrode.

There are several advantages to using sEEG or ECoG over scalp EEG and MEG. While scalp-level recordings can reflect the combination of multiple underlying cortical signals, direct recordings at the cortical surface with ECoG or within the cortex with sEEG provide significantly better spatial resolution. Perhaps more importantly, direct recordings of patches of tissue do not rely on statistical estimates of sources, providing higher spatial certainty. While local field potentials have been found to reflect the activity of 3-5 mm and 50 to 200 µm of cortex for macro- and microelectrode contacts respectively (Buzsáki, 2004; Harris et al., 2016; Kreiman et al., 2006), the exact spatial resolution of intracranial signals depends on multiple factors including electrode size, spacing, impedance, referencing and frequency of interest, and investigated signal frequency. Another advantage over scalp EEG is an improved signal-to-noise ratio by one to two orders of magnitude (Ball et al., 2009; Ramantani et al., 2016), which allows for the detection of more subtle effects. Furthermore, while scattering and artifactual activity in scalp signals limit the analysis of high frequency components (Cohen, 2014; Pfurtscheller and Cooper, 1975), intracranial signals capture brain oscillations in the high gamma range that have been associated with local neuronal firings (Lachaux et al., 2012; Nir et al., 2007; Ray & Maunsell, 2011). Finally, the contacts along the electrodes can be used to directly stimulate the brain, making it possible to modulate brain activity intracranially and study the effect of neuromodulation on consciousness (for a review, see Raccah et al., 2021).

It is important to acknowledge that intracranial research is not devoid of limitations (Parvizi & Kastner, 2018). First, it does not provide homogeneous sampling of the brain across participants. Implantation schemes are individualized based on suspected epileptic sources and brain anatomy. This limits the types of inferences that can be drawn from group statistics, although methods exist to alleviate this challenge (Mentzelopoulos et al., 2024; Mercier et al., 2022). It also underscores the importance to report results based on channels' MNI coordinates and individual anatomy, and not just based on labelled regions of interest. Second, the sparse coverage resulting from the clinical constraints of human intracranial research imply that some areas of the brain may not be sampled from at all. As a result, intracranial studies might fail to record from important neural populations and do not have as broad a coverage as scalp EEG and MEG. Note, however, that implants can cover a large portion of the cortex, enabling the study of multiple brain regions in a single experiment, although occipital and parietal regions are often undersampled. Third, because recordings are obtained from patient populations that have pathological brain activity patterns, care should be taken when interpreting the results and their generalizability. Recorded signals can be contaminated by epileptic activity, and neuroplasticity can occur following cortical damage caused by epileptic seizures (Parvizi & Kastner, 2018). These concerns are mitigated by the removal of epileptic artifacts and showing an effect of interest in multiple patients with different suspected epileptogenic networks (Mercier et al., 2022). Despite the inherent challenges of interpreting intracranial signals, this method has the unique advantages listed above, and can greatly contribute to our understanding of the NCC. Importantly, though, even the most precise recordings cannot replace careful behavioral designs and paradigm innovations in distinguishing NCCs from their prerequisites and consequences. In the following sections, we describe the results of studies that have investigated the NCC using human intracranial data.

## Population-level cortical correlates of visual consciousness

Macroelectrode measurements (i.e., ECoG or sEEG) capture the collective activity of large populations of neurons and can sample large portions of the brain in a single participant. Such local field potential data are particularly relevant to NCC research, as many theoretical predictions suggest that consciousness emerges from neural population activity (Albantakis et al., 2023; Brown et al., 2019; Lamme, 2018; Malach, 2021; Mashour et al., 2020). For a summary of the studies that used macroelectrode measurements to study the visual NCC, see Table 1.

We start in the posterior visual system, where the occipitotemporal cortex provides the earliest and most consistent intracranial responses to visual stimuli. Several sEEG studies focused on event-related potentials in response to perceived and not perceived visual word stimuli in backward masking paradigms (Gaillard et al., 2006a; 2006b; 2009; Naccache et al., 2005). In one representative study, the authors found an early response in channels located in the occipital and temporal lobes, regardless of whether the word was masked or unmasked (Gaillard et al., 2009). Later responses (200 ms after stimulus onset) observed in the same channels were stronger in the unmasked condition, indicating a possible role of these responses in perceptual consciousness. Another study examined the low gamma response (30 to 70 Hz) from ECoG channels primarily located in the occipitotemporal cortex while participants performed a backward masking task, in which they were asked to recognize different categories of visual stimuli (Fisch et al., 2009). Channels that responded specifically to the presentation of one stimulus category, often to faces, exhibited an early (starting 150 ms post-stimulus), nonlinear response following recognized vs. unrecognized stimuli. These channels were located mainly in non-retinotopic high-order visual cortical regions, including the lateral occipital cortex and fusiform gyrus. Altogether, these early human intracranial studies indicate that early-latency visual processing steps, reflected in broadband and low gamma activity, occur irrespective of whether a stimulus is consciously perceived or not. They also identified a candidate NCC: later activity in the occipitotemporal region responsible for higher-order visual processing.

More recent human intracranial studies investigating the NCC focused mainly on High Gamma band Activity (HGA). HGA is believed to be more focal than broadband activity, reflecting the firing rates of the population of neurons located near depth electrodes or beneath strip or grid electrodes (Lachaux et al., 2012; Nir et al., 2007; Ray & Maunsell, 2011) or dendritic processes, depending on the depth at which it is recorded (Leszczynski et al., 2020). One study examined HGA from ECoG grids placed on the lateral occipitotemporal cortex and the fusiform gyrus while participants reported the visibility of near-threshold pictures (Aru et al., 2012b). In some trials, they were presented with the picture for the first time, but in other trials, they had been previously exposed to the same, non-degraded picture. Manipulations of physical evidence and prior exposure produced the same effect of increasing reported visibility. However, while six out of seven channels with category-specific HGA exhibited an effect of physical evidence, none demonstrated an effect of prior exposure. Since both manipulations influenced reported visibility but prior exposure did not influence HGA in the lateral occipitotemporal cortex and the fusiform gyrus, the authors concluded that these regions are not part of the NCC. A subsequent sEEG study utilized a similar manipulation, examining the impact of repetition suppression (where neural activity is reduced through repeated stimulus exposure) on HGA in the ventral visual cortex (not only in the fusiform gyrus) in response to masked and unmasked words (Vidal et al., 2014). Repetition suppression influenced the reduction of response amplitude in two out of three word-responsive channels in both masked and unmasked conditions. The results of these two studies revealed a complex pattern: on the one hand, HGA in the lateral occipitotemporal cortex and the ventral visual cortex correlated with stimulus strength. On the other hand, it also correlated with another factor that does not appear to play a role in visibility (repetition suppression), and did not correlate with a non-sensory factor that affects visibility reports (prior exposure). These results raise the possibility that activity in occipitotemporal cortex regions reflecting higher-order visual processing may be a precursor to the NCC. Another possibility is that prior exposure influenced *reports* of visibility, and not perceptual consciousness itself, in which case HGA in occipitotemporal cortex may still reflect an NCC proper.

Other intracranial studies provided evidence that went in the other direction, suggesting that under certain experimental conditions, HGA in the occipitotemporal cortex might still be related to visibility per se. They focused on neural responses to face stimuli, because these responses in the occipitotemporal cortex are well-characterized and robust (Kanwisher et al., 1997; Rossion et al., 2023). A recent analysis investigated HGA in a task where participants provided immediate detection responses to degraded, near-threshold face stimuli (Li et al., 2024). Consistent with previous findings (Fisch et al., 2009), the response of face-selective ECoG channels located over the ventral visual cortex was stronger for faces reported as seen than for unseen faces. This effect was more pronounced in channels located in the more anterior part of the ventral visual cortex. The amplitude and peak times of HGA correlated with response accuracy and latency across channels, as would be expected from a response associated with perceptual consciousness. In another ECoG study, channels located on the surface of the lateral and ventral temporal cortex were used. A face presented in one interval and a blank in another were both masked with continuous flash suppression (Baroni et al., 2017). Participants performed a 2-interval forced choice paradigm, where they reported which of the two intervals contained the face. They then rated its visibility on the perceptual awareness scale, a four-point scale that has been found to offer a more appropriate measure of perceptual consciousness than binary scales (Ramsøy & Overgaard, 2004; Sandberg et al., 2010). HGA captured by channels in the fusiform gyrus correlated better with subjective visibility than with actual stimulus intensity, suggesting that this response plays a crucial role in the emergence of conscious face percepts. A re-analysis of those data complemented by a backward masking task in the same patients found similarity between conscious contents and measures of information integration based on neural activity in occipitotemporal cortex (Haun et al., 2017). Altogether, these studies found that HGA in the ventral visual cortex evoked by face stimuli correlates with visual consciousness of faces in several ways. It is possible that this signal consists of a content-specific NCC involved in the subjective experience of faces, while content-specific responses for other categories of experiences have been less thoroughly investigated.

So far, we have focused on the occipitotemporal cortex, leaving aside the role of frontoparietal activations in perceptual consciousness. We now turn to studies focusing on more anterior correlates of consciousness. A seminal study examined intracranial equivalents to the P3 component, a late positivity observed over frontoparietal electrodes, in response to clearly visible oddball letter stimuli or to subliminal stimuli (Brázdil et al., 2001). They identified a P3 component in both conditions, which exhibited an earlier peak and shorter duration in the subliminal condition. Additionally, this “subliminal P3” was less anatomically widespread than its supraliminal counterpart, showing no detectable response in the dorsolateral prefrontal cortex. Prefrontal channels in backward masking studies also showed a stronger evoked response for unmasked vs. masked stimuli (Gaillard et al., 2009) and an increased gamma response for recognized vs. unrecognized stimuli (Fisch et al., 2009). However, these studies could not assess if the differences between the two conditions reflected a NCC or activity linked to preparing a report.

To mitigate the report confound, two recent sEEG studies used an elaborate contrastive task. Gratings were presented at different intensities around the perceptual threshold, but participants could only prepare their responses after a delay of 650 ms. In the first study, the prefrontal cortex exhibited stronger broadband activity and HGA in response to consciously perceived than non-consciously perceived gratings, regardless of stimulus contrast (Fang et al., 2024a). In the second study, the authors further demonstrated that the onset times of the broadband response in the lateral prefrontal cortex followed a bimodal distribution, in line with the possibility that the first cluster relates to perceptual consciousness while the latter cluster relates to post-perceptual processing (Fang et al., 2024b, Figure 3A). This interpretation assigns a role to the lateral prefrontal cortex in perceptual consciousness. It should be treated with caution, though, because it does not depend on an analysis that directly assesses the influence of post-perceptual confounds beyond report preparation.

Beyond studying the responses of channels or regions in isolation, several studies have looked at functional connectivity to assess how interactions within or between brain regions shape perceptual consciousness. Phase synchrony analyses revealed that long-range coherence in the beta band between pairs of channels across the cortex increased in the unmasked vs. masked condition (Gaillard et al., 2009). In a subsequent study, several markers of functional connectivity across prefrontal channels were also found to increase significantly when participants consciously saw the gratings (Fang et al., 2024a). As discussed in detail below, intralaminar thalamic activity was later found to drive this prefrontal activity (Fang et al., 2025). Because none of these studies controlled for report-related activity using no-report paradigms, it remains unclear to what extent these connectivity patterns reflect perceptual consciousness or post-perceptual processing. Additionally, it is unknown how the signals associated with perceptual consciousness in the occipitotemporal cortex interact with frontoparietal regions.

Other human intracranial studies have benefited from more extensive cortical coverage, and could look at the role of large-scale brain networks in visual consciousness. One study suppressed images with a variant of continuous flash suppression where images are presented long enough to “break” into consciousness (Shan et al., 2022). The authors could classify whether low-frequency oscillatory activity (4-30 Hz) preceded or followed images’ entry into awareness from all recorded channels. While the results were interpreted in light of perceptual consciousness, they might also reflect the mechanisms linked to the perceptual alternations involved in breaking continuous flash suppression (Stein & Peelen, 2021). Another study with broad cortical coverage examined the responses of brain networks during a delayed detection response to near-threshold face stimuli (Herman et al., 2019). The authors conducted analyses that separated channels across clusters based on the HGA following stimuli reported as seen. The first cluster, which included the primary visual cortex, posterior parietal lobe, inferior lateral frontal cortex, and orbitofrontal cortex, was associated with initial activation for both perceived and unperceived stimuli, followed by a dip in activity below baseline, and finally by a late reactivation (>600 ms after stimulus onset) in response to perceived stimuli only. A second cluster, linked to the default mode network, was deactivated starting 300 ms after stimulus onset. Lastly, the ventral visual cortex and frontoparietal association cortices showed an early (starting <200 ms after stimulus onset) and later sustained wave of activation. In another study where participants performed a detection task on near-threshold gratings (Liu et al., 2023), five distinct clusters, including channels from the same four lobes, were found to exhibit stronger activations to stimuli reported as seen vs. unseen. However, the results should be interpreted cautiously, because they do not distinguish between the NCC and its consequences, such as the neural activity associated with reporting one's perceptual consciousness.

Few studies have attempted to pair contrastive tasks that do not require participants to provide a report on a visual stimulus with intracranial recordings. In one study, participants were presented with a ring whose contrast changed several times at fixed intervals during each trial (Vidal et al., 2015). In the middle of the trial, the ring was always presented at the same contrast, but adaptation suppression made it such that it was either visible when preceded by a lower contrast ring or invisible when preceded by a higher contrast ring. In the main experiment, participants were not required to provide any detection response about the ring at that point in the trial, but were asked to report a ring offset two seconds later. A stronger increase in low-frequency oscillatory activity (8-24 Hz) was observed in response to visible vs. invisible rings in all brain lobes. More surprisingly, HGA increased in the invisible ring condition and decreased in the visible ring condition in contacts located in the occipital, temporal, and frontal lobes, including the fusiform gyrus, anterior insula, and inferior frontal gyrus. One explanation for this result is that the task set introduced by reporting a disappearance at the end of the trial led to the encoding of a saliency signal in the invisible condition. Note that it is difficult to make conclusions about the NCC from this study alone, as the visible and invisible conditions also differed in terms of the contrast of the preceding ring.

We recently performed a sEEG study where cross-task decoding was used to disentangle the NCC from post-perceptual confounds linked to perceptual reports (Stockart et al., 2025). Participants performed three experiments on the same sequence of near-threshold face stimuli. Multivariate decoders were trained on HGA to discriminate between seen and unseen stimuli in an immediate-response experiment, where participants were tasked to press on a button as soon as they saw a face. The same decoders could predict whether a face stimulus was reported as seen or unseen in all regions of interest when tested on HGA in a second, delayed-response experiment. In the third, no-report experiment, participants were presented with faces at two intensities around perceptual threshold, with the assumption that faces at the higher intensity would be perceived more often. Stimulus intensity was classified above chance-level performance in the ventral visual cortex by decoders trained on detection reports in the first experiment, but not in the parietal and frontal cortices. These results indicate that HGA activity in response to faces in the ventral visual cortex are not a post-perceptual confound (Figure 3B).

Altogether, studies that investigated the cortical correlates of visual consciousness point to a role of neural responses starting ~250 ms after stimulus onset in the non-primary visual cortex and prefrontal cortex (although the precise timing of these responses may depend on the task and across trials; Pereira et al., 2022; Salti et al., 2019). Lateral and ventral occipitotemporal signals are observed in the absence of report preparation, but studies have found contradictory results as to whether they correlate better with perceptual consciousness or physical features of the stimuli. More work is also required to establish the extent to which early responses in contrastive tasks in the prefrontal cortex, and particularly its lateral part, reflect report preparation.

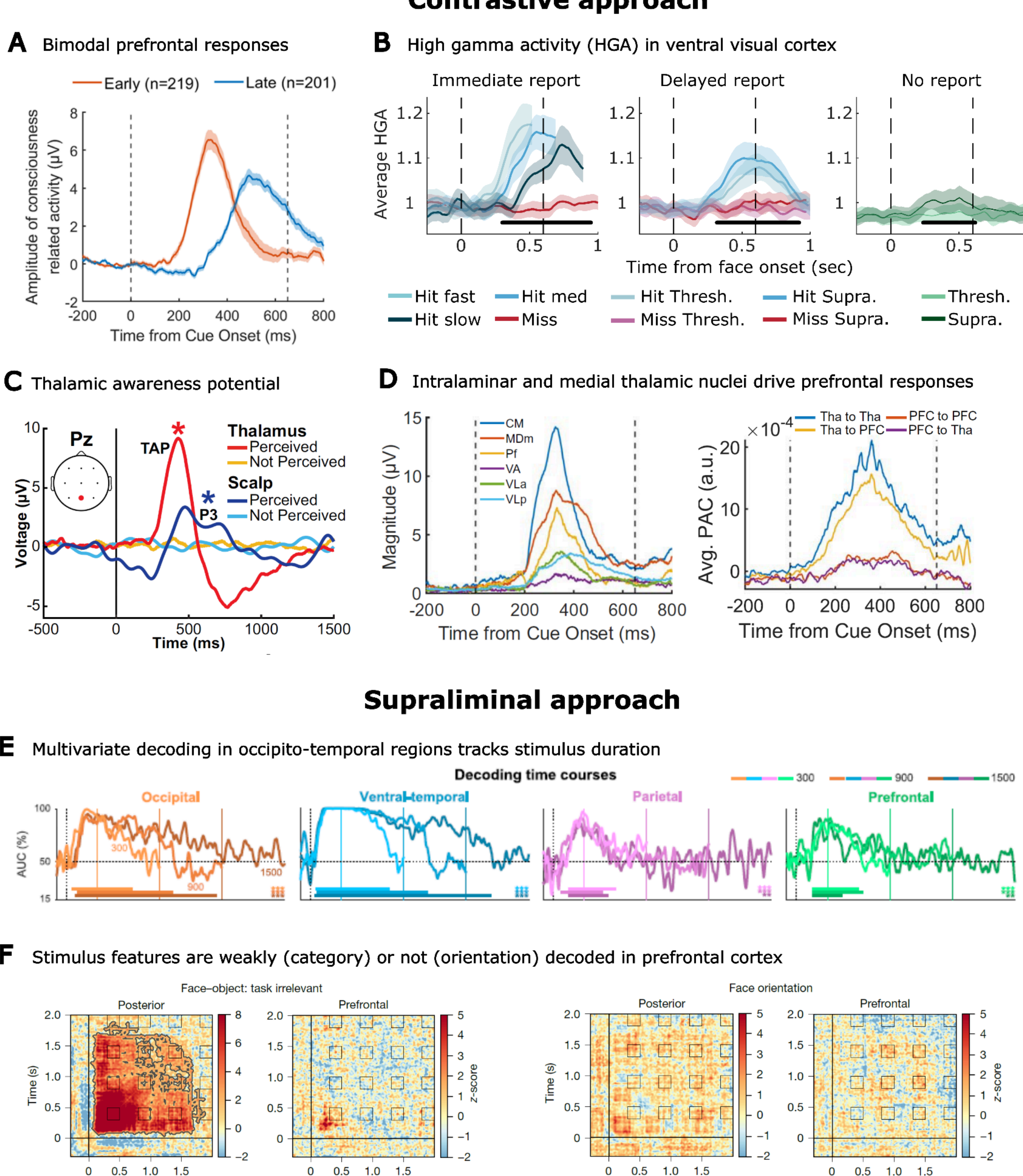

*Figure 3.* Summary of key results of intracranial studies of perceptual consciousness. *A.* Broadband activity in response to seen vs. unseen face stimuli in the prefrontal cortex follows a bimodal distribution, leading the authors to speculate that early responses may be associated with perceptual consciousness. Panel adapted from Fang et al. (2024b) (CC BY-NC 4.0). *B.* In the ventral visual cortex including the face fusiform area, HGA increases track face stimulus detection in immediate and delayed report experiments, and stimulus intensity in a no-report experiment. Panel adapted from Stockart et al. (2025) (CC BY-NC-ND 4.0). *C.* A thalamic awareness potential is observed in response to seen, but not unseen, face

stimuli. Panel adapted from Kronemer et al. (2022) (CC BY 4.0). *D.* Medial and intralaminar thalamic nuclei respond more to perceived face stimuli than ventral nuclei (left panel) and drive activity in the prefrontal cortex as shown by phase-amplitude coupling (PAC; right panel). Adapted from Fang et al. (2025). *E.* Multivariate decoding in the occipital and ventral-temporal cortices, but not the parietal and prefrontal cortices, tracks stimulus duration. Panel adapted from Vishne et al., 2023 (CC BY 4.0). *F.* Stimulus category can be decoded for the entire duration of the stimulus in the posterior cortex but not in the prefrontal cortex, while stimulus orientation decoding does not track the duration of the stimulus in either region of interest. Panel adapted from Cogitate Consortium et al. (2025) (CC BY 4.0).

| Study | Implant type | Coverage | Sample size | Stimuli | Suppression method | Task type | Main result |
|---|---|---|---|---|---|---|---|
| Brazdil et al. (2001) | sEEG | Temporal and frontal | 13 | Letters | Near-threshold (oddball task) | Report | The subliminal P3 peaks earlier and is shorter in duration than the supraliminal P3. |
| Naccache et al. (2005) | sEEG | Amygdala | 3 | Words | Backward masking | Report | Masked word emotion affected broadband activity ~870 ms after stimulus onset. |
| Gaillard et al. (2006a) | sEEG | Occipito-temporal | 7 | Words | Backward masking | Report | Repetition of previously masked words exerts a long-lasting effect on neural signals. |
| Gaillard et al. (2006b) | sEEG | Occipito-temporal | 1 | Words | Backward masking | Report | Cortical event-related response to masked words was confined to ventral pathways and not as sustained as for masked words. |
| Gaillard et al. (2009) | sEEG | All lobes (mostly posterior) | 10 | Words | Backward masking | Report | Perceptual consciousness is associated with sustained activity in the prefrontal cortex and increased long-distance synchrony. |
| Fisch et al. (2009) | ECoG | All lobes (mostly posterior) | 11 | Categories | Backward masking | Report | Stimuli reported as seen induce a nonlinear increase of activity in higher-order visual cortex. |
| Aru et al. (2012b) | ECoG | Occipito-temporal | 6 | Pictures | Near-threshold | Report | A manipulation of visibility by prior exposure indicates that the higher-order visual cortex is not a NCC. |
| Vidal et al. (2014) | sEEG | Occipito-temporal (ventral) | 3 | Words | Backward masking | Report | Repetition suppression in the ventral visual cortex occurs for perceived and non-perceived stimuli. |
| Vidal et al. (2015) | sEEG | All lobes | 9 | Shape (ring) | Contrast adaptation | No-report | Stimulus disappearance triggered decreases in low-frequency activity and increases in HGA. |
| Baroni et al. (2017) | ECoG | Occipito-temporal | 5 | Faces | Near-threshold + Continuous flash suppression | Report | Decoding of activity in ventral visual cortex and lateral visual cortex correlates with subjective visibility better than with stimulus strength. |
| Haun et al. (2017) | ECoG | Occipito-temporal | 6 | Faces | Near-threshold + Continuous flash suppression + Backward masking | Report | Conscious contents show correspondence to integrated information measures based on neural activity. |

| Herman et al. (2019) | ECoG + sEEG | All lobes | 9 | Faces | Near-threshold | Report | Perceived stimuli, but not non-perceived stimuli, are accompanied by large-scale. network switching. |
|---|---|---|---|---|---|---|---|
| Kroenemer et al. (2022) | sEEG | Thalamus | 7 | Faces | Near-threshold | Report | A thalamic event-related potential with onset ~250 ms is observed in response to perceived vs. non-perceived stimuli. |
| Shan et al. (2022) | sEEG | All lobes | 7 | Not reported | Breaking continuous flash suppression | Report | Activity from every channel could be used to discriminate between brain activity before and after stimuli broke from suppression. |
| Liu et al. (2023) | sEEG | Occipital, temporal, parietal and frontal | 13 | Gratings | Near-threshold | Report | Clustering analyses showed an interaction between exogenous attention and conscious report. |
| Li et al. (2024) | ECoG | Ventral temporal | 4 | Faces | Near-threshold | Report | Face-specific activity in the ventral visual cortex is stronger and more reliable for perceived than non-perceived stimuli. |
| Fang et al. (2024a) | sEEG | Mostly prefrontal | 6 | Gratings | Near-threshold | Report | Stimuli reported as perceived induce prefrontal activations independently from motor preparation. |
| Fang et al. (2024b) | sEEG | Mostly prefrontal | 9 | Gratings | Near-threshold | Report | The onset of broadband activity in response to perceived vs. non-perceived stimuli in the lateral prefrontal cortex follows a bimodal distribution. |
| Fang et al. (2025) | sEEG | Thalamus and prefrontal | 5 | Gratings | Near-threshold | Report | Thalamic responses to perceived vs. non-perceived stimuli are earliest and strongest in medial and intralaminar nuclei, and drive lateral prefrontal cortex activity. |
| Stockart et al. (2025) | sEEG | All lobes | 29 | Faces | Near-threshold | Report + no-report | A neural code in the ventral visual cortex reflects (1) stimulus detection and (2) stimulus intensity in a no-report experiment. |

*Table 1*. Summary of ECOG and sEEG studies using the contrastive approach to study visual NCCs in humans.

## Population-level cortical correlates beyond vision

Another key question in the study of consciousness is the extent to which the putative NCCs found with visual stimuli extend to other sensory modalities (Sanchez et al., 2020). The intracranial studies discussed so far, like the whole field of consciousness, mostly focus on the visual modality (Faivre et al., 2017). One study, which measured sEEG correlates during contralateral near-threshold median nerve stimulation, found evidence for a key role of the parietal operculum in tactile consciousness (Albertini et al., 2025). HGA in this region, as well as in the superior parietal, motor, and premotor cortices, was modulated by stimulus intensity in both report and no-report experiments. Tonic responses in the parietal operculum, however, best distinguished between stimulations reported as perceived vs. not perceived and showed the earliest inflection points in their neurometric curve, indicative of all-or-none responses.

Macroelectrode recordings were also used to examine the auditory NCC. In a case study, one patient implanted with several ECoG grids completed a tone repetition detection task on informationally masked stimuli (Dykstra et al., 2016), a method that can render salient sounds imperceptible (Gutschalk et al., 2008). Event-related responses to masked targets reported as detected vs. undetected over Heschl's gyrus and the inferior frontal cortex exhibited an early negativity in a window spanning from 100 to 200 ms after stimulus onset. Late positive responses were also observed more broadly across temporal and frontal regions. Interestingly, HGA showed earlier, more focal increases than the broadband response. A recent, large-scale study with extensive anatomical coverage investigated the auditory NCC with ECoG and sEEG (Christison-Lagay et al., 2025). HGA increases in response to perceived stimuli were recorded from the right caudal middle frontal gyrus as early as 25 ms after stimulus onset, and then propagated to the auditory cortex, as well as to insular, frontal, and occipital regions. HGA increases in response to non-perceived stimuli remained mostly confined to the auditory cortex. The difference in HGA response to perceived vs. non-perceived stimuli was broadly similar to what was observed in vision with a comparable design (Herman et al., 2019), except for a sustained activation of the auditory cortex.

Overall, studies of the NCC beyond the visual modality offer leads for shared intracranial neural responses to perceptual consciousness across sensory modalities (e.g., frontoparietal cortex; Table 2). Like many studies on the NCC in the visual modality, however, several of them did not control for post-perceptual processing. As a result, it is a possibility that shared activations across sensory modalities reflect the consequences of perceptual consciousness rather than a NCC.

| Study | Implant type | Coverage | Sample size | Stimuli | Suppression method | Task type | Main result |
|---|---|---|---|---|---|---|---|
| Dykstra et al. (2016) | ECoG | Inferior temporal and inferior frontal | 1 | Tones (auditory) | Informational masking | Report | Heschl's gyrus and the inferior frontal cortex increase their activity for heard vs. unheard sounds. |
| Christison-Lagay et al. | ECoG + sEEG | All lobes | 31 | Categories (auditory) | Near-threshold | Report | The cortical and subcortical networks involved in auditory consciousness are similar to those |

| (2025) | | | | | | | in visual perception. |
|---|---|---|---|---|---|---|---|
| Albertini et al. (2025) | sEEG | All lobes | 30 | Median nerve stimulation (tactile) | Near-threshold | Report + no-report | Tonic HGA responses in the parietal operculum (1) are all-or-none, (2) differentiate between stimuli reported as perceived and not perceived and (3) are observed in a no-report task. |

*Table 2*. Summary of ECOG and sEEG studies using the contrastive approach to study non-visual NCC in humans.

## Single-neuron cortical correlates of perceptual consciousness

Macroelectrodes measure the activity of large populations of neurons, and thus do not capture finer-grained activity patterns. On the other hand, microelectrodes can capture single-unit or multi-unit activity reflecting the firing rates of single or small groups of neurons[3]. Though microelectrodes recordings generally yield a few units per participant restricted to specific brain regions, they provide a unique opportunity to track single neurons firing and its relations to perceptual consciousness (Table 3). This constitutes an interesting piece of the puzzle in uncovering the NCC, for several reasons. First, single-unit activity provides the level of granularity necessary to dissociate neurons representing the percept from neighboring neurons involved in task-related confounds (e.g., motor preparation or arousal). Percepts represented by sparse coding involving a small, specific population might only be identified with microelectrode recordings. Second, key computations at the level of population dynamics can only be uncovered by unit-level recordings (Vyas et al., 2021). Animal studies reveal that various cognitive processes are encoded within neuronal subspaces that only emerge when single-unit activity is analyzed as lower-dimensional projections of the broader neural activity manifold (Mante et al., 2013; Ebitz & Hayden, 2021; Jayazeri & Afraz, 2017). Third, while many prominent theories predict that perceptual consciousness is a system-level property, there is still a lack of clarity about the appropriate level at which consciousness should be studied (Aru et al., 2020; Whyte et al., 2024). For instance, micro-scale measurements are required to test the prediction of Dendritic Integration Theory (Bachmann et al., 2020) that the integration of feedforward and feedback signals occurs at the level of individual pyramidal neurons. Finally, beyond spatial granularity, single-unit activity also provides excellent temporal granularity, which is crucial for testing theories that rely on the precise timing of spikes (e.g., neural synchrony). Only microelectrode recordings can confirm whether individual neurons lock their spikes to a specific phase, a mechanism hypothesized for binding features into a conscious whole (Singer & Gray, 1995).

[3] Unlike macroelectrode recordings, which are primarily used for clinical purposes, microelectrode recordings - whether through microfilaments added to sEEG clinical electrodes or through multielectrode arrays - are currently used solely for research purposes.

Microelectrode recordings used in NCC research were mostly obtained from microwires at the tip of sEEG electrodes in the medial temporal lobe, including contacts in the amygdala, entorhinal cortex, hippocampus, and parahippocampal gyrus. In those studies, participants were presented with visual stimuli from different categories (e.g., faces, houses, tools, etc.), which they either perceived consciously or not. A first study had participants perform a flash suppression paradigm, where a stimulus initially presented to both eyes is suppressed by the sudden presentation of a different stimulus in the dominant eye (Kreiman et al., 2002). Medial temporal neurons that responded specifically to one stimulus or one stimulus category did so when the stimulus was consciously perceived but not when it was not perceived. Another study presented participants with unmasked vs. backward masked visual stimuli, finding stronger responses for unmasked compared to masked stimuli (Quiroga et al., 2008). This study also confirmed that neuronal responses in the medial temporal lobe were specific to certain types of perceptual contents.

While studies investigating human single-unit activity in the medial temporal cortex uncovered stronger neuronal activity for perceived over non-perceived stimuli, there is also evidence for partially conserved processing when the stimuli were not consciously perceived. This effect was particularly strong in a more recent study, in which stimuli that were pre-screened for eliciting preferential neuronal responses were presented in an attentional blink paradigm (Reber et al., 2017). Relatively preserved activity for stimuli reported as non-consciously perceived was measured from many medial temporal neurons, including several hundred milliseconds after stimulus onset. This response was stronger for posterior than anterior contacts. It was also generally less robust and delayed compared to the response to stimuli that were reported as consciously perceived. These findings led the authors to argue that the NCC consists of precise spiking patterns and not a binary activation status of neurons. It is also possible that the degree to which unconscious processing occurs in the medial temporal lobe depends on the blinding method used, with attentional manipulations leaving more processing intact (Kanai et al., 2010).

We know of only two studies that investigated single neuron responses to visual consciousness outside of the medial temporal lobe. A recent study used microelectrode arrays to measure single and multi unit activity from the surface of the human lateral occipital cortex in response to different categories of visual stimuli (Vanhoyland et al., 2025). In backward masking and flash suppression experiments, recorded neurons were found to respond more strongly to stimuli reported as perceived than to stimuli reported as not perceived, similar to what was observed in the medial temporal lobe. However, the stimulus category could be decoded both when the stimuli were reported as perceived and not perceived. Interestingly, the latency and activation profiles of the neuronal response varied greatly with the precise location of the implants on the lateral occipital cortex. In a third, binocular rivalry experiment, stimulus category could be decoded from firing rates as early as 1.5 s before participants reported a perceptual transition to that stimulus. This result is consistent with another study investigating single neuron and multi-unit responses in the medial temporal lobe during binocular rivalry (Gelbard-Sagiv et al., 2018). In both studies, firing rates preceded the change in perceptual consciousness, indicating the existence of unconscious representations among occipitotemporal neurons. In the second study, some participants also had implants in the pre-supplementary motor area and anterior cingulate cortex, and the activity from neurons in these regions could predict the perceptual transitions even earlier than neurons in the medial temporal lobe. This result suggests a role of these structures in resolving the conflict between the two rivaling images, and therefore a contribution as a precursor for consciousness. It remains unknown, however, whether decoding of perceived contents based on firing rates in the lateral occipital cortex precedes or comes after activity in the frontal cortex.

Moreover, none of these studies controlled for the consequences of perceptual consciousness, leaving it unclear whether the results reflect NCCs or post-perceptual processing. This concern was mitigated in a study that used near-threshold vibrotactile stimuli in both report and no-report experiments (Pereira et al., 2021). One human participant was implanted with an ECoG grid and a microelectrode array in the posterior parietal cortex. The researchers identified neurons that responded more strongly when the stimulus was consciously perceived. Motor and decisional confounds were controlled for by demonstrating that neuronal responses could also be modulated by stimulus intensity in a no-report version of the task. These results accordingly suggest that neuronal responses in the posterior parietal cortex are unlikely to reflect the consequences of the NCC and may play a role as neural correlates of tactile consciousness. They also corroborate the results of a sEEG study that found a key role for the parietal lobe in tactile consciousness (Albertini et al., 2025), albeit in a different part of that lobe (parietal operculum).

| Study | Implant type | Coverage | Sample size | Stimuli | Suppression method | Task type | Main result |
|---|---|---|---|---|---|---|---|
| Kreiman et al. (2002) | Micro-wires | Medial temporal | 14 | Categories | Flash suppression | Report | The activity of neurons in medial temporal lobe follows reported percepts. |
| Quiroga et al. (2008) | Micro-wires | Medial temporal | 5 | Categories | Backward masking | Report | The activity of neurons in medial temporal lobe follows reported percepts. |
| Reber et al. (2017) | Micro-wires | Medial temporal | 21 | Pre-screened stimuli | Attentional blink | Report | The strength and timing of neurons' responses to seen vs. unseen stimuli follows a posterior to anterior gradient in medial temporal lobe. |
| Gelbard-Sagiv et al. (2018) | Micro-wires | Medial temporal and frontal | 9 | Pre-screened stimuli | Binocular rivalry | Report | Changes in firing rates of medial temporal lobe neurons reflect the preferred stimulus start > 1 s before a perceptual transition. |
| Pereira et al. (2021) | ECoG + array | Posterior parietal | 1 | Vibrations (tactile) | Near-threshold | Report + no-report | Neurons in the posterior parietal cortex reflect evidence accumulation, confidence and stimulus intensity in the absence of reports. |
| Pereira et al. (2025) | Microelectrodes | Subthalamic nucleus + thalamus | 32 | Vibrations (tactile) | Near-threshold | Report | Thalamic and subthalamic nuclei differentiate between perceived vs. non-perceived vibrotactile stimuli. |
| Vanhoyland et al. (2025) | Array | Lateral occipital cortex | 4 | Categories | Backward masking + Flash suppression + Binocular rivalry | Report | Classification of stimulus category based on neural population responses only occurs for stimuli reported as seen. |

*Table 3*. Summary of studies using the contrastive approach to study single-neuron NCC in humans.

## Subcortical correlates of perceptual consciousness

So far, this review has solely focused on NCC candidates located in the cortex. Subcortical structures, particularly the thalamus, have also been implicated in consciousness (Bachmann et al., 2020; Blumenfeld, 2023; Llinás et al., 1998; Whyte et al., 2024). While it has long been believed that subcortical networks are crucial in the slow modulations of arousal and attention, and that thalamo-cortical interactions significantly contribute to consciousness (Llinás et al., 1998), investigations of subcortical responses in contrastive tasks are still relatively rare (e.g., Christison-Lagay et al., 2025; Fang et al., 2025; Haegens et al., 2014; Kronemer et al., 2022; Levinson et al., 2021; Pereira et al., 2025; Tauste Campo et al., 2019).

To test the proposal that thalamic nuclei play a key role in enabling perceptual consciousness, a study directly measured local field potentials in the intralaminar thalamus of human participants during a contrastive task on near-threshold face stimuli (Kronemer et al., 2022). The authors uncovered a biphasic event-related potential, the thalamic awareness potential, in response to faces reported as seen vs. unseen (Figure 3C). Concurrent scalp EEG data showed that the thalamic awareness potential occurred at an intermediate latency between an early (the visual awareness negativity) and a late (P3b) EEG component. Another recent study jointly investigated thalamic and prefrontal responses to near-threshold grating stimuli in a task where participants could not prepare their reports during a delay period (Fang et al., 2025). Changes in voltage were found for gratings reported as seen vs. unseen in all recorded thalamic nuclei, reliably across trials and as early as 200 ms post-stimulus. They appeared across more channels, showed earlier onsets, and were stronger in magnitude in intralaminar and medial nuclei than in ventral nuclei. Functional connectivity analyses revealed that the flow of information following stimulus presentation was from these thalamic nuclei to the lateral prefrontal cortex, rather than in the reverse direction, which led the authors to suggest that the medial and intralaminar thalamic nuclei act as a gate to perceptual consciousness (Figure 3D).

Thalamic responses to perceptual consciousness were not only found in response to visual stimuli. This thalamic awareness potential was observed in the right thalamus of one patient in response to auditory stimuli reported as heard vs. unheard (Christison-Lagay et al., 2025; Kronemer et al., 2022). Another study that examined intraoperative recordings of neurons located in the subthalamic and thalamic nuclei during a near-threshold vibrotactile task identified neurons with differential responses between trials where participants reported feeling vs. not feeling the stimuli (Pereira et al., 2025).

Together, the findings of these studies suggest that the thalamus is not only involved in perceptual consciousness by serving as a sensory relay and modulating long-term levels of consciousness. Rather, the medial and intralaminar nuclei appear to act as a gate for stimuli to access perceptual consciousness and further processing in the cortex (Fang et al., 2025, Kronemer et al., 2022). More research is required to determine whether subcortical responses like the thalamic awareness potential play a direct role in the NCC or in the precursors or consequences of perceptual consciousness.

## The supraliminal approach

No-report conditions are not the only major development of the last decade in the study of the NCC. Although the field is still mainly focused on the contrastive approach, other paradigms have been proposed and used (Lepauvre & Melloni, 2021). The “supraliminal approach”, which consists in presenting clearly visible stimuli to participants, is increasingly popular (Cogitate Consortium et al., 2025; Gerber et al., 2017; Noy et al., 2015). Participants are required to perform a task that keeps them attentive, but requires little cognitive demand. Such tasks include pressing a button when a subset of stimuli is presented (the brain response to the other stimuli is then the main focus of analysis), reporting when a stimulus is repeated or recalling the stimuli at the end of the experiment. The results are often interpreted as being less confounded by post-perceptual activity than in contrastive tasks, where participants have to perform a demanding task on near-threshold stimuli. The two main advantages of the supraliminal approach are that (1) it is credible that participants consciously perceive the stimuli in most if not all trials even when they do not directly report on those stimuli and (2) suprathreshold stimuli evoke stronger neural responses. One important pitfall, however, is that it cannot distinguish between unconscious and conscious processing. Because stimuli are always suprathreshold, the neural activity that occurs only when participants are conscious of the stimuli cannot be isolated. Thus, studies using the supraliminal approach cannot be regarded as a means to isolate and detect NCCs. However, their results can affect our assessments of NCCs. For example, if an NCC candidate previously detected using the contrastive approach is not observed in response to suprathreshold stimuli, this casts strong doubts on it being an NCC. In addition, presenting stimuli at full contrast avoids the drawback of working with stimuli presented near the perceptual threshold, making it easier to study the dynamics of perceptual consciousness (e.g., the maintenance of a conscious percept).

In that regard, an ECoG study investigated how well HGA tracked stimulus duration in the occipitotemporal cortex by presenting full-contrast face and house stimuli for variable durations (Gerber et al., 2017). Duration-tracking followed a posterior-to-anterior gradient, with many duration-tracking channels in the early visual cortex and very few in the inferior temporal cortex. Category-selectivity showed the opposite pattern and was found to be strongest in the temporal cortex. Intriguingly, the only two channels encoding category and track duration were located in the posterior fusiform face area, consistent with contrastive studies indicating that this brain area plays a special role in conscious perception of faces. A re-analysis of data from this study including channels in the parietal cortex and prefrontal cortex found that stimulus contents could be decoded for the whole time that the stimulus was presented in occipital and ventral temporal cortices, but only for a short time after stimulus onset in frontal and parietal regions (Vishne et al., 2023; Figure 3E). Another human intracranial study presented consistent results, finding decoding of stimulus contents in occipitotemporal regions for the entire duration of 1.5 second long stimuli (Broday-Dvir et al., 2023). Interestingly, while decoding of the stimulus category (faces vs. places) was successful for face-selective contacts in ventral temporal and lateral occipital cortices, only exemplars and not categories could be decoded in the early visual cortex.

In a recent large-scale adversarial collaboration study, different categories of clearly visible stimuli were paired with intracranial recordings, amongst other imaging modalities (Cogitate Consortium et al., 2025). Two regions of interest were specified to test predictions of the global neuronal workspace and integrated information theories of consciousness: the prefrontal cortex and a posterior “hot zone” including occipital, temporal, and parietal regions. While decoding of stimulus categories was found in both regions of interest, it was very short-lived in the prefrontal cortex and did not track the duration of the stimulus (Figure 3F), consistent with prior studies (Broday-Dvir et al., 2023; Vishne et al., 2023). Together, the results of these three studies indicate that populations of neurons in the occipitotemporal cortex are involved in maintaining conscious percepts over time. At the anatomical level, these results extend a previous ECoG study with large-scale cortical coverage that used the supraliminal approach to assess whether visual NCCs are found in frontoparietal or higher-order sensory cortices (Noy et al., 2015). HGA increases in response to all stimuli and to specific stimulus categories in the higher-order visual cortex had a larger amplitude, but not earlier latencies, than in frontal and parietal cortices. The results were interpreted as failing to clearly arbitrate between a stronger role of the higher visual or frontoparietal cortex in perceptual consciousness.

Finally, another research group used the supraliminal approach to study HGA in response to visual linguistic stimuli. In a first study, they used a block design where letter stimuli were presented at 1 Hz in an active session, and no stimuli were presented in a passive session (Li et al., 2019). While both the visual cortex (including V1 and the fusiform gyrus) and the default mode network showed sustained inhibition during the active session, only the former transiently increased its activity in response to the stimulus. A re-analysis of these data showed widespread and early transient brain responses to the letter stimuli, starting in the occipital and frontal cortices, and then propagating to the fusiform, medial temporal, and frontal cortices (Khalaf et al., 2023). These results were confirmed with word stimuli in a study that leveraged data from a large human intracranial database, demonstrating very early signals in the frontal cortex potentially involved in signal detection (Kwon et al., 2021). They highlight the importance of studying network dynamics besides the response of isolated regions.

Studies using the supraliminal approach are summarized in **Table 4**. Altogether, they make it easier to study facets of consciousness like maintenance without relying on reports. However, the lack of an unconscious condition makes it particularly hard to distinguish NCCs from unconscious activations that precede it. Thus, their insights are best evaluated in combination with studies that used the contrastive approach.

| Study | Implant type | Coverage | Sample size | Stimuli | Main result |
|---|---|---|---|---|---|
| Noy et al. (2015) | ECoG | All lobes | 43 | Categories | Content-specific, high-magnitude visual cortex signals are followed by content-invariant frontoparietal signals. |
| Gerber et al. (2017) | ECoG | Occipito-temporal | 10 | Faces and houses | The duration for which a stimulus is presented is better reflected by activity in posterior than anterior visual cortex. |
| Li et al. (2019) | ECoG + sEEG | All lobes | 11 | Letters | Presentation of conscious stimuli leads to both sustained and transient dynamics in cortical network activity. |
| Kwon et al. (2021) | ECoG + sEEG | All lobes | 158 | Words | Visual, medial temporal and frontal cortices form a signal detection network involved at the onset of perceptual consciousness. |
| Khalaf et al. (2023) | ECoG + sEEG | All lobes | 11 | Letters | A large range of regions from all cortical lobes are involved in an early detection network. |
| Broday-Dvir et al. (2023) | ECoG + sEEG | All lobes | 13 | Faces and places | Sustained representation of conscious stimuli depends on similarity distances between activation patterns in VVC. |
| Vishne et al. (2023) | ECoG | All lobes | 10 | Faces and houses | The population response in occipitotemporal regions encodes stimuli's temporal dynamics. |
| Cogitate Consortium et al. (2023) | ECoG + sEEG | All lobes | 34 | Categories | Stimulus category can be decoded for the entire stimulus duration in the posterior region of interest, and after stimulus onset in the prefrontal region of interest. |

*Table 4.* Summary of intracranial studies using the supraliminal approach to study the NCC in humans.

## Insights gained from intracranial studies on theoretical accounts of consciousness

When taken together, what do these results teach us about theoretical accounts of consciousness? A large number of such accounts have been proposed (for reviews, see Doerig et al., 2021; Kuhn, 2024; Seth & Bayne, 2022). They differ in many aspects, which are not always easy or possible to arbitrate empirically (Evers et al., 2024; Fazekas et al., 2024; Mudrik et al., 2025; Northoff & Lamme, 2020; Yaron et al., 2022). Two avenues for evaluating theories of consciousness against each other are their anatomical and temporal predictions about the NCC. Intracranial recordings are well suited for doing so, given their joint anatomical and temporal resolution. They also lend themselves to predictions about the mechanism that subtends perceptual consciousness. In what follows, we discuss the anatomical and temporal insights brought by intracranial consciousness research.

*Anatomical insights*

Localist theories predict that the NCC is to be found in (non-primary) sensory cortices. For example, the recurrent processing theory holds that perceptual consciousness corresponds to feedback activity across these areas (Lamme, 2018; for a similar account, see also Malach, 2021). The integrated information theory predicts that perceptual consciousness relies on a "posterior hot zone" reflecting maximal integrated information in the brain and spanning the temporal, parietal, and occipital cortices (Boly et al., 2017). The results reviewed here are broadly consistent with the NCC being located in non-primary sensory cortices. Specifically, lateral and ventral occipitotemporal activity has been shown to correlate with visual consciousness, including in no-report conditions (Fisch et al., 2009; Stockart et al., 2025; Vanhoyland et al., 2025). A suggested dissociation is that non-primary sensory regions play a role as precursors for consciousness rather than reflecting a NCC proper (Melloni et al., 2011). This possibility is supported by the finding of one sEEG study that HGA tracks one source of evidence (stimulus strength) but not another (previous exposure), both of which affect visibility reports (Aru et al., 2012b). An alternative explanation, however, is that prior exposure only influences *reports* of visibility, and not perceptual consciousness itself, in which case the NCC should only reflect the effect of stimulus strength and not prior exposure. Further investigation of the precursors of the NCC will be necessary to settle this issue. Notably though, the critical claim of localist theories is that posterior activations are sufficient for consciousness. Thus, finding positive evidence for the involvement of posterior areas in perceptual consciousness does not, by itself, confirm these theories. For that, one has to show that no additional activations (e.g., in prefrontal areas) are required, or rely on neurological lesion data showing that patients who are missing one part of the cortex have intact perceptual consciousness (but interpretations of such findings can be contentious, Boly et al., 2017; Odegaard et al., 2017).

Indeed, this is the main contrast between localist theories and cognitive ones, which assign a crucial role to the prefrontal cortex in perceptual consciousness. The global neuronal workspace theory holds that a given content is consciously experienced when a non-linear "ignition" broadcasts this content throughout the brain via thalamocortical loops and pyramidal neurons in parietal and prefrontal cortices (Mashour et al., 2020). Higher-order theories hold that a first-order representation of a perceptual content in sensory cortices is insufficient for this content to become conscious, and that a higher-order representation/indexing, generally predicted to be in the prefrontal cortex, is needed (Brown et al., 2019; Lau & Rosenthal, 2011). Although none of the contrastive studies reviewed here directly investigated the role of the prefrontal cortex for ignition or higher-order mechanisms, some of them are broadly consistent with cognitive theories, since activity in the prefrontal cortex, and specifically in its lateral part, correlated with reportedly perceived content (Brazdil et al., 2001; Fang et al., 2024a; 2024b; Gaillard et al., 2009; Gelbard-Sagiv et al., 2018). This is partly supported by the finding of an onset response in the prefrontal cortex for clearly visible, task-irrelevant stimuli, and the ability to decode their content from the prefrontal cortex, though not for all aspects (Cogitate Consortium et al., 2025).

In a no-report task, we recently failed to find a neural code reflecting perceptual consciousness in the prefrontal cortex and posterior parietal cortex (Stockart et al., 2025). However, it is important to note that this null result is not evidence that these regions are not part of the NCC. A single-neuron study found NCC-like neuronal activity in the posterior parietal cortex (Pereira et al., 2021), suggesting that macroelectrode recordings may not have the required level of granularity to pick out the relevant activity (Naccache et al., 2021). Also, although studies using fMRI and MEG indicate that some activity patterns linked to the NCC reflect post-perceptual processing, some prefrontal activity is still observed in no-report conditions (Hatamimajoumerd et al., 2022; Kronemer et al., 2022; Sergent et al., 2021). Though this criticism is relevant to most intracranial studies, it does not apply to no-report ones, where post-perceptual effects are less expected (though see Block, 2019). Notably, given the mixed pattern of results found in no-report intracranial experiments (e.g., Albertini et al., 2025; Cogitate Consortium et al., 2025; Stockart et al., 2025; Vishne et al., 2023), more research is needed to elucidate this point and assess the strength of intracranial evidence supporting cognitive theories.

We further note that not all cognitive theories hold that conscious contents themselves should be represented in the prefrontal cortex. Given that content representation occurs in sensory cortices, prefrontal cortex activity does not need to reflect that content directly but can simply serve as a “pointer” to the relevant sensory representations (Block, 2024; Lau, 2022).

*Temporal insights*

The temporal dynamics of putative NCCs can also be used to assess theories of consciousness. For instance, several localist theories predict that the NCC should be observed for the entire duration for which a stimulus is presented, while the global neuronal workspace predicts that it should only be observed at stimulus onset and offset (Cogitate Consortium et al., 2025; Malach, 2021; Melloni et al., 2021). The visual studies reviewed here support the idea that sustained activity in ventral temporal regions tracks the time course of visual consciousness for suprathreshold stimuli. In contrast, frontal regions generally exhibit more transient activity locked on stimulus onset, with only anecdotal evidence for offset responses (Cogitate Consortium et al., 2025). While some exceptions have been identified using the contrastive method (Fang et al., 2024a), these might be related to the working memory processes necessary for stimulus report. Overall, these findings are consistent with the idea that conscious percepts are primarily encoded in posterior regions, whereas the frontal cortex might provide brief, top-down signals that facilitate gating or selecting percepts represented in posterior regions. This interpretation aligns with several theoretical frameworks put forward to explain perceptual consciousness. Some higher-order theories of consciousness suggest that prefrontal regions play a key role in determining which sensory representations reach consciousness (Fleming, 2020; Lau, 2019). It has also been proposed that global neuronal workspace neurons provide top-down compressed information to lower-level cortical areas (Mashour et al., 2020). Top-down activity also plays a central role in recurrent processing and predictive processing theories, even if such activity need not originate in the prefrontal cortex (Seth & Bayne, 2022). The observation that posterior regions display sustained NCCs while prefrontal activity tends to be transient is also consistent with findings in non-human primates. These works show that the inferior temporal and prefrontal cortices are both involved in higher-order visual processing of faces (Kornblith & Tsao, 2017), though the latter is more related to behaviorally-relevant categorization (Freedman et al., 2003). Whether the putative top-down activity observed in the prefrontal cortex reported here reflects post-processing alone or plays a more direct role in consciousness remains to be established (Panagiotaropoulos, 2024).

The high temporal resolution of electrophysiological data makes it possible to investigate the latencies of putative NCCs. In the non-primary visual cortex, they have been found to range from <100 to >300 ms following stimulus onset (Stockart et al., 2025; Vanhoyland et al., 2025), although one consistent finding is that activity representing conscious contents follows an initial wave of unconscious processes (Fisch et al., 2009). Variable latencies as early as 200 ms and as late as over 500 ms after stimulus onset have also been observed in the prefrontal and posterior parietal cortices (Fang et al., 2024b; Pereira et al., 2021). The observation of such latency differences within the same brain region appears to depend on several factors including the specific neural populations being recorded (Fang et al., 2024b), reaction times (Li et al., 2024; Pereira et al., 2021; Stockart et al., 2025) and the experimental paradigm used (e.g., near-threshold stimuli seem to lead to later latencies than backward masked stimuli). This suggests that perceptual consciousness is a dynamic phenomenon and that NCCs might occur at variable timings (He, 2023; Pereira et al., 2022; Salti et al., 2019; Sergent, 2018). Therefore, predictions about the exact timing of raw latencies probably have limited use in comparing theories of consciousness.

However, the high temporal resolution of intracranial recordings can be leveraged in other ways. Paired with no-report paradigms, it can provide unique insights into the temporal dynamics of putative NCCs across behavioral paradigms. It also makes it possible to analyse the frequency content of neural responses and to determine the direction of flow of information between brain regions with connectivity analyses (Cogitate Consortium et al., 2025; Fang et al., 2025). For example, recent research indicates that the intralaminar thalamus drives activity in the lateral prefrontal cortex and not the other way around (Fang et al., 2025), consistent with theoretical predictions of the global neuronal workspace and others (Blumenfeld, 2023; Mashour et al., 2020). Precise predictions about how percepts are maintained in the brain and expected interareal connectivity can be readily assessed with intracranial recordings.

## Limitations

We reviewed human intracranial research on perceptual consciousness. These studies are informative about candidate NCC regions, but are also subject to pitfalls, some of which are discussed below.

First, because human intracranial NCC research in modalities other than vision is still in its infancy, it is not clear whether existing findings reflect mechanisms that are shared across consciousness of different modalities, or only to the visual one (Sanchez et al., 2020; note that this problem pertains to NCC research in general, and not only to intracranial studies). More research about the brain regions that subtend content-invariant NCCs (if any) is required, and one way to achieve this is to look at brain responses across sensory modalities. Second, cortico-centric bias and scarce coverage of subcortical structures make it unclear what role subcortical structures play in perceptual consciousness, and whether they are constitutive of the NCC. A growing body of research is starting to uncover the role of the thalamus in perceptual consciousness (Kronemer et al., 2022; Fang et al., 2025; Pereira et al., 2025).

Third, the large heterogeneity in definitions of regions of interest makes it hard to compare the results of different studies and take full advantage of the fine anatomical resolution of intracranial recordings. A useful recommendation for nuanced anatomical interpretations and data synthesis across studies would be to make MNI coordinates of relevant contacts readily available. Fourth, it remains unclear whether the findings of many studies using the contrastive approach indeed reflect the NCC or associated confounds. Only a few of the human microelectrode recording studies used the same parameters for the unconscious and the conscious conditions (Pereira et al., 2021; Reber et al., 2017). This raises the concern that some of the findings reflect the physical differences between the conditions, rather than changes in consciousness itself.

Fifth, some concerns have been raised about the contrastive approach and no-report paradigms. No-report experiments do not fully address confounding activity from post-perceptual cognitive processing (Block, 2019; 2024). Covert decisions or mind-wandering are likely to co-occur with perceptual consciousness, even when participants are not engaged in a task. A related concern involves attentional confounds when comparing the results of report and no-report experiments: While participants are incentivized to pay attention in report experiments because of the task, they are likely to be less attentive in no-report experiments (Lau, 2025). Consequently, they may perceive stimuli less often in no-report than in report experiments, leading to reduced statistical power in detecting NCCs. Note that these first two issues also apply to the supraliminal approach.

Sixth, there are reasons to be wary of interpreting participants' reports of not having perceived a stimulus as a ground truth (Fahrenfort et al., 2025; Schmidt, 2015; for a recent discussion of best practices and pitfalls in studying unconscious processes, see Stockart, Schreiber et al., 2025). The setting of a decision criterion affects participants' categorization of trials as conscious or unconscious, independently of phenomenology (Macmillan, 1986). Also, researchers should be mindful that participants' reports of being conscious of something do not necessarily imply being conscious of the task-relevant features of the stimulus (Michel, 2023). Although these concerns expose some weaknesses, we believe that none are fatal to the study of the NCC with the contrastive approach (a position also shared by others; Doerig et al., 2021).

However, they highlight that it is unlikely that there is a “perfect experiment” that addresses all sources of concern simultaneously.

Lastly, it is essential to remember that the best correlate is still just that: a correlate. Therefore, it is important to recognize that understanding consciousness will necessitate an evaluation of its mechanisms. This involves integrating intracranial recordings with a computational approach, alongside a falsifiable theoretical framework that makes precise empirical predictions. One previous study has compared behavior and electrophysiological signals with simulations from a computational model of evidence accumulation (Stockart et al., 2025). Other fields have also benefited from mechanistic approaches through the combination of biologically-constrained recursive neural networks with (animal) electrophysiology. In perceptual decision-making for instance, reverse engineering these recursive neural networks helped reveal how task-relevant features of a stimulus are selected (Barbosa et al., 2023; Mante et al., 2013) or how top-down signals affect choice signals in sensory regions (Wimmer et al., 2015). Nonetheless, such computational approaches are still seldom used in human studies of perceptual consciousness.

## Perspectives

*Towards improved behavioral paradigms*

The inherent quality of intracranial data in no way diminishes the necessity to combine them with meticulously controlled behavioural paradigms. It will be important for future invasive and noninvasive studies to more systematically distinguish the NCC from its precursors and consequences. Future intracranial research should consider the use of manipulations to isolate NCCs from their precursors, such as participants’ expectations and attention (Aru, et al., 2012b; Melloni et al., 2011; Wyart et al., 2012). There is also value in directly characterizing activity that is known to come before perceptual consciousness and how it interacts with later NCC-related activity. For instance, intracranial studies have shown that pre-stimulus activity in the amygdala and ventral striatum is predictive of whether stimuli are perceived or not (Guex et al., 2023; Slagter et al., 2017). Non-veridical percepts, where participants report perceiving a stimulus when none was presented, can also be used to distinguish neural activity associated with the NCC from that associated with stimulus processing (Filimonov et al., 2025; Haarsma et al., 2023; Stockart et al., 2025).

Other studies should focus on distilling the NCC from its consequences. Few intracranial studies used the contrastive approach in combination with a no-report paradigm, where participants passively perceive stimuli, to study a NCC not confounded by decisional processes (Albertini et al., 2025; Pereira et al., 2021; Stockart et al., 2025). Combined with electrophysiology studies in non-human primates (de Lafuente & Romo, 2006; Hesse & Tsao, 2020; Kapoor et al., 2022; Panagiotaropoulos et al., 2012) and fMRI studies in humans (Hatamimajoumerd et al., 2022; Kronemer et al., 2022), they raise the interesting possibility that many cortical areas, including the visual cortex, posterior parietal cortex and lateral prefrontal cortex, could be constitutive of the NCC. Beyond no-report paradigms, the contrastive approach could be paired with other experimental manipulations that address post-perceptual confounds. One group recorded early brain responses free of motor confounds by making the response contingent on a cue presented after a delay period (Fang et al., 2024a; 2024b; 2025). But because it still requires a report, this strategy does not control for brain areas that are recruited in the decision-making process (Aru et al., 2012a; De Graaf et al., 2012; Sergent & Naccache, 2012). The observed responses could be some of the neural generators of the P3b, a scalp event-related potential that is not observed when decisional activity is accounted for (Cohen et al., 2020). Alternative controls include the decoding of task-irrelevant features (Cogitate Consortium et al., 2025; Mante et al., 2013).

Yet another interesting avenue is to investigate the dynamics of perceptual consciousness of stimuli at perceptual threshold, and rely on paradigms in which participants are asked to reproduce subjective aspects of perception such as its duration (Msheik et al., 2025). Considering these temporal aspects of perceptual consciousness could help isolate the true NCC. A promising direction for research would involve combining supraliminal and contrastive tasks within the same study. With cross-task decoding, researchers could for instance determine if the same neural code is involved in perceptual consciousness in the contrastive task and percept duration tracking in the supraliminal task.

Beyond traditional NCCs, the excellent sensitivity of recordings made directly in contact with the cortex could make it possible to carry out very short experiments involving very few or even a single trial. This would make it possible to identify NCCs without any learning in so-called “zero-shot experiments”. These experiments have been seminal in the study of consciousness, particularly in elucidating the links between attention and consciousness (e.g., inattentional blindness; Rock et al., 1992), but are difficult to characterise at the neural level using non-invasive measures because the phenomenon cannot be repeated many times.

*Towards multiscale recordings*

An important question that arises from NCC research with intracranial recordings is the level of granularity at which the NCC should be studied. While most studies focus on isolated brain regions, it is likely that only adopting a meso-scale approach would miss a part of the explanation. On the one hand, the NCC is likely to be constituted by the coordinated activity of whole-brain networks and not by activity in anatomical regions taken in isolation (Blumenfeld, 2023). For example, it is meaningful to consider deactivation of the default mode network as a whole in response to perceptual consciousness (Herman et al., 2019; Kwon et al., 2021). As a result, more studies with whole-brain coverage could consider pooling channels across cortical networks instead of arbitrarily large cortical regions that sometimes include entire lobes (Herman et al., 2019; Liu et al., 2023). On the other hand, it is important to understand the micro-scale with investigations of single or multi-unit firing rates. These micro-recordings can provide information that is invisible when recording large populations of neurons.

*Towards a within-patient research of contents and levels of consciousness*

For clinical reasons, patients with intracranial electrodes sleep during naps and nights in the hospital and undergo general anesthesia before the electrodes are explanted. Although methodologically complex and potentially disturbing for patients, recording intracranial signals under these conditions, along with a contrastive study during wakefulness, provides an exceptional framework for jointly studying the contents and levels of consciousness (Eichenlaub et al., 2020; Jiang et al., 2017; Krom et al., 2020). A particularly interesting suggestion is the use of multivariate decoding of contents as a way to assess participants' levels of consciousness, thus individualizing analyses to participants' specific neuroanatomy and processing times (Fischer et al., 2025). Such analyses have been performed on scalp EEG (King et al., 2013), but the spatiotemporal resolution of intracranial recordings could provide better sensitivity. The suprathreshold approach's non-reliance on reports makes it suitable to identify brain responses that differ across levels of consciousness. Pairing this approach with intracranial recordings in sleep has already shown that HGA increases in response to sounds are largely preserved in the auditory cortex, including in the deeper stages of sleep (Hayat et al., 2022). Stimulus-evoked event-related desynchronization was also strongly reduced during sleep (Hayat et al., 2022), indicative of disrupted feedback connections (van Kerkoerle et al., 2014). Similarly, the primary auditory cortex was found to respond strongly to auditory click trains during anesthesia, while the response of the surrounding cortex was disrupted (Krom et al., 2020). Beyond sleep and anesthesia, this approach could also be paired with investigations of epileptic seizures as an altered state of consciousness to be studied in its own right (Arthuis et al., 2009; Blumenfeld et al., 2004; Bonini et al., 2016; Englot et al., 2010; Lambert et al., 2012; Mateos et al., 2018). In the context of NCC research, investigations of conscious contents across levels of consciousness could be used to provide indications that some candidate NCCs actually reflect unconscious, sensory processing.

*Towards other applications of intracranial interventions*

We have only scratched the surface of the research possibilities offered by intracranial recordings in human patients. While studying NCCs using the contrastive approach remains a standard in consciousness research, complementary approaches specific to intracranial recordings merit discussion. First, the preoperative assessment of the epileptogenic zone often requires clinicians to perform electrical stimulation, typically by applying a weak electric current between two contacts on selected electrodes. These clinical stimulations often evoke conscious experiences (e.g., visual hallucinations, somatosensory sensations, etc.) that can be analysed to evaluate the causal role of a specific cortical region in a given conscious percept (Raccah et al., 2021). With the right clinical setup, electrical stimulations can also be triggered at specific timings relative to the presentation of a stimulus, to study the role of brain regions or networks at specific latencies (e.g., Hampson et al., 2018). Additionally, electrical stimulations produce signals that travel along axonal fibers and can be measured by remote electrodes. These measurements, known as cortico-cortical evoked potentials, provide accurate information about functional connectivity (Lemaréchal et al., 2022) and could enhance our understanding of the circuitry of the NCCs, considered as a network rather than isolated regions. The rapid technological development of electrodes that enable the measurement of a large number of neurons (e.g., neuropixels; Chung et al., 2022; Coughlin et al., 2023; Leonard et al., 2023) or the assessment of signals previously little studied in humans (e.g., voltammetry; Kishida et al., 2016) suggests an improved characterization of electrophysiological and electrochemical correlates of consciousness in the near future.

Human intracranial research is also expanding beyond studying the brain of patients who are implanted with electrodes in the context of pre-surgical assessments. New types of implants with different clinical targets are increasingly used. Chronic deep brain and cortical implants are used to stimulate and record cortical and subcortical structures in patients with a wide range of psychiatric and neurological conditions, including disorders of consciousness (Cao et al., 2024), migraines (Leone, 2006), obsessive compulsive disorders (Chabardes et al., 2020), depression (Hitti et al., 2021), Parkinson's disease (Deuschl et al., 2006), tumors (Boussen et al., 2016) and epilepsy (Gummadavelli et al., 2015). For example, a recent study found that stimulation of the subthalamic nucleus of patients with obsessive compulsive disorders did not affect their perception of near-threshold stimuli (Kist et al., 2024). The emerging diversity in invasive electrophysiological recordings offers new avenues of research for studying perceptual consciousness. Micro-electrode recordings outside of the medial temporal lobe made it possible to investigate the contrastive response of neurons in other regions of the human brain (Pereira et al., 2021; 2025; Vanhoyland et al., 2025). Additionally, implants in non-epileptic patient populations can target structures that are typically not sampled or undersampled in epileptic participants' population, but could play a role in perceptual consciousness (e.g., thalamus; Fang et al., 2025).

## Concluding remarks

Despite decades of research, the neural bases of perceptual consciousness remain contentious. Human intracranial recordings, with their excellent joint temporal and anatomical resolution, already provide key insights into the neural correlates of consciousness that complement those provided by noninvasive methods. In the coming years, improved paradigms, multiscale recordings, diverse implants, and new analyses promise to shed mechanistic insights on how the brain builds subjective experience from the surrounding world.

Acknowledgments: Funded by the European Union. Views and opinions expressed are however those of the author(s) only and do not necessarily reflect those of the European Union or the European Research Council Executive Agency. Neither the European Union nor the granting authority can be held responsible for them. This work is supported by an ERC grant (Volta, 101125379). FS is supported by a post-doctoral study grant of the Fyssen Foundation.